\documentstyle[emulateapj,epsf]{article}

\def\simless{\mathbin{\lower 3pt\hbox
   {$\rlap{\raise 5pt\hbox{$\char'074$}}\mathchar"7218$}}} %< or of order
\def\simgreat{\mathbin{\lower 3pt\hbox
   {$\rlap{\raise 5pt\hbox{$\char'076$}}\mathchar"7218$}}} %> or of order
\def\vvec{{\mbox{\boldmath $v$}}}
\def\rvec{{\mbox{\boldmath $r$}}}
\def\Vvec{{\mbox{\boldmath $V$}}}
\def\etavec{{\mbox{\boldmath $\eta$}}}

\def\xhat{{\mbox{\boldmath $\hat x$}}}
\def\yhat{{\mbox{\boldmath $\hat y$}}}%
\def\ehat{{\mbox{\boldmath $\hat e$}}}
\def\zhat{{\mbox{\boldmath $\hat z$}}}
\def\rvec{{\mbox{\boldmath $r$}}}
\def\wvec{{\mbox{\boldmath $w$}}}
\def\ehatone{{\mbox{\boldmath $\hat e_1$}}}
\def\ehattwo{{\mbox{\boldmath $\hat e_2$}}}
\def\ehatthree{{\mbox{\boldmath $\hat e_3$}}}
\def\Jvec{{\mbox{\boldmath $J$}}}
\def\Avec{{\mbox{\boldmath $A$}}}
\def\Ahat{{\mbox{\boldmath $\hat A$}}}
\def\uvec{{\mbox{\boldmath $u$}}}
\def\Jhat{{\mbox{\boldmath $\hat J$}}}
\def\What{{\mbox{\boldmath $\hat w$}}}

\def\crossprod{{\mbox{\boldmath $\times$}}}
\def\dotprod{{\mbox{\boldmath $\cdot$}}}
\def\kms{{\rm\,km\,s^{-1}}}

\def\msun{M_\odot}

\def\be{\begin{equation}}
\def\ee{\end{equation}}
\def\baray{\begin{eqnarray}}
\def\earay{\end{eqnarray}}
\def\min{{\rm min}}
\def\tid{{\rm tid}}

\def\simless{\mathbin{\lower 3pt\hbox
   {$\rlap{\raise 5pt\hbox{$\char'074$}}\mathchar"7218$}}} %< or of order
\def\simgreat{\mathbin{\lower 3pt\hbox
   {$\rlap{\raise 5pt\hbox{$\char'076$}}\mathchar"7218$}}} %> or of order
\def\vvec{{\mbox{\boldmath $v$}}}

\def\rvec{{\mbox{\boldmath $r$}}}

\def\Vvec{{\mbox{\boldmath $V$}}}
\def\Jvec{{\mbox{\boldmath $J$}}}

\def\yhat{{\mbox{\boldmath $\hat y$}}}
\def\xhat{{\mbox{\boldmath $\hat x$}}}
\def\zhat{{\mbox{\boldmath $\hat z$}}}
\def\mmc{{\rm{mmc}}}

\def\kms{{\rm\,km\,s^{-1}}}

\def\gcm3{{\rm g\,\, cm^{-3}}}
\def\msun{M_\odot}

\def\be{\begin{equation}}
\def\ee{\end{equation}}
\def\baray{\begin{eqnarray}}
\def\earay{\end{eqnarray}}

\begin{document}

\title{Formation of an Evanescent Proto-Neutron Star 
Binary and the Origin of Pulsar Kicks}

\author{Monica Colpi\altaffilmark{1} and Ira Wasserman\altaffilmark{2}}

\altaffiltext{1}{Dipartimento di Fisica G. Occhialini, Universit\`a di
Milano Bicocca, Piazza della Scienza 3. I-20126 Milano, Italy}
\altaffiltext{2}{Center for Radiophysics and Space Research, 
Cornell University, Ithaca, NY 14853}
\authoremail{colpi@mib.infn.it}

\begin{abstract}
If core collapse leads to the formation of a rapidly rotating
bar-unstable proto-neutron star surrounded by fall-back material, then
we might expect it to cool and fragment to form a double
(proto)-neutron star binary into a super-close orbit.  The lighter
star should survive for awhile, until tidal mass loss propels it
toward the minimum stable mass of a (proto)-neutron star, whereupon it
explodes.  Imshennik \& Popov have shown that the explosion of the
unstable, cold star can result in a large recoil velocity of the
remaining neutron star. Here, we consider several factors that
mitigate the effect and broaden the range of final recoil speeds, in
particular the finite velocity and gravitational deflection of the
ejecta, a range of original masses for the low mass companion and its
cooling history, rotational phase averaging of the momentum impulse
from non-instantaneous mass loss, and the possibility of a common
envelope phase. In spite of these mitigating factors, we argue that
this mechanism can still lead to substantial neutron star recoil
speeds, close to, or even above, $1000\kms$.

\end{abstract}


\section{Introduction} 
It is recognized that radio pulsars have peculiar space velocities
between $\approx 30 \kms$ and $\approx 1600\kms$, significantly greater than
those of their progenitor stars. 
The highest speed ever recorded is from
the Guitar Nebula pulsar (Cordes et al. 1993; Cordes \& Chernoff 1998),   
while the lowest has been recently measured for B2016+28
by Brisken  et al. (2002) 
from very accurate VLBA  pulsar parallaxes.
Statistical studies, aimed at inferring
the peculiar velocity at birth from the observed speed, give
mean three-dimensional velocities of 100 -- 500 $\kms$ for the
isolated pulsars (Lyne \& Lorimer 1994; Lorimer, Bailes, \&
Harrison 1997; Hansen \& Phinney 1997; Cordes \& Chernoff 1998;  
Arzoumanian, Chernoff, \& Cordes 2002).

An early explanation for the
large space velocities called for {\it recoil} in a close
binary that becomes unbound at the time of (symmetric) supernova explosion 
(Blaauw 1961; Iben \& Tutukov
1996). Now, a number of observations  hint at a {\it natal}
origin of these high space velocities, or at a combination of orbital
disruption 
(when the progenitor
lives in a binary) and internal kick reaction.
Evidence for a kick at the time
of neutron star birth is now found in a variety
of systems: in runaway O/B associations (Leonard \& Dewey 1992),
in highly eccentric Be/NS binaries (van den Heuvel \& Rappaport 1986;
Portgies-Zwart \& Verbunt 1996),
in the binary pulsar J0045-7319 (Kaspi et al. 1996; to explain 
its current spin-orbit configuration), and in double neutron
star binaries such as B1913+16 (Bailes 1988; Weisberg,
Romani, \& Taylor 1989; Cordes, Wasserman, \& Blaskiewicz 1990; 
Kramer 1998; Wex, Kalogera \& Kramer 2000)
where  
misalignment between the spin and orbital 
angular momentum axes indicates velocity asymmetry
in the last supernova.
All these observations support the view that the formation of neutron stars
is accompanied by anisotropic explosion.  
This notion is 
bolstered by evolutionary studies of
binary populations (Dewey \& Cordes 1987; Fryer \& Kalogera
1997; Fryer, Burrows, \& Benz 1998)
and by studies (e.g. Cordes \& Wasserman  1984) 
on the survival of binary systems into
their  late evolutionary stages after supernova explosion.

Among the physical processes that have been proposed to account
for the kicks are large-scale 
density asymmetries seeded in the pre-supernova core (leading to
anisotropic shock propagation), asymmetric  neutrino emission
in presence of ultra-strong magnetic fields 
(see Lai, Chernoff, \& Cordes 2001 for
a review), or off-centered electromagnetic dipole emission from the
young pulsar (Harrison \& Tademaru 1975).
However, none of these  mechanisms can explain  kicks 
as large as $\sim 1600 \kms$.

In this paper we reconsider the idea first put forward by Imshennik \&
Popov (1998) that, in the collapse of a rotating core, one or more 
self-gravitating lumps of neutronized matter may form in close orbit
around the central nascent neutron star, transfer mass 
in the short lived binary, and ultimately
explode, 
causing the remaining, massive neutron star to acquire a
substantial kick velocity, as high as the highest observed.
The light member explodes as mass transfer drives it below
the minimum stable mass for a neutron stars.
In the light star, stability is lost upon decompression  
by the $\beta-$decaying  neutrons and nuclear fissions by radio-active
neutron-rich nuclei 
(Colpi, Shapiro \& Teukolsky 1989, 1991; Blinnikov et al. 1990), 
that deposit energy driving
matter into rapid expansion 
(Colpi, Shapiro \& Teukolsky 1993; Sumiyoshi et al. 1998).  
The kick tries its origin from the orbital motion of
this evanescent super-close
binary, that forms in the collapse
of a rapidly rotating (isolated) iron core.

We will study several effects that may
modify the magnitude of the kick, such as gravitational bending
of the exploding debris, rotational averaging of the momentum impulse, 
orbit decay, and delayed neutron star cooling.  

Formation of a proto-neutron star companion around 
the main neutron star has never been verified 
in numerical simulations 
due to computational limitations.
For this reason we elaborate on a study of
Bonnell (1994) on the formation of binary/multiple systems in collapsing
gas cloud cores and its extension to the stellar core collapse in the
aftermath of a supernova explosion (Bonnell \& Pringle 1995)
to motivate our working hypothesis. The 
formation of such exotic binaries  
has been conjectured to occur in many works
(Ruffini \& Wheeler 1971; Clark \& Eardley 1977;
Blinnikov, Novikov, Perevodchikova, \& Polnarev 1984; Nakamura \&
Fukugita 1989; Stella \& Treves 1987).

\section {Light fragments around proto-neutron stars}

\subsection {The Scenario}

Formation of a light companion around
a main body implies breaking of spherical and axial symmetry during
collapse and following core bounce.  During dynamical collapse, unstable bar
modes ($m=2$) can grow in a fluid (even non rotating) that may end
with fragmentation. However this is known to 
occur only if the cloud core contracts almost isothermally, as in the
case of star's formation from unstable cold gas clouds (Bonnell 1994).  Core 
collapse in type II supernovae is far from isothermal (it is
described by a effective polytropic index $\gamma\simeq 1.3$) so that
instabilities of this type do not have time to grow (Lai 2000)
\footnote{Large-scale asymmetries imprinted in the iron core prior to
collapse may lead to anisotropic explosions that produce kicks 
as indicated by Goldreich, Lai \& Sahrling (1996). Whether they lead
to fragmentation is unknown.}, and  
simulations of non-axisymmetric rotating core collapse 
confirm this trend (Rampp, Muller \&
Ruffer 1998; Centrella, New, Lowe, \& Brown, 2001).
Can fragmentation/fission be excited after core bounce ?

Rapid rotation in equilibrium bodies is known to excite non-axisymmetric
dynamical instabilities and these instabilities may grow in the
proto-neutron star core. Interestingly, core collapse
simulations of unstable rotating iron cores (Heger, Langer \& Woosley
2000; Fryer \& Heger 2000) or polytropes (Zwerger \& Muller 1997)
indicate that proto-neutron stars, soon after formation, can rotate
differentially above the dynamical stability limit
set when the rotational to gravitational potential energy ratio
$T_{\rm rot}/\vert W\vert$ is larger than the value $\beta_{\rm dyn}=0.25-0.26$ 
(Saijo et
al. 2001).  
Strong non-linear
growth of the dominant bar-like deformation ($m=2$) is seen in
these cores (described as polytropes by Rampp,
Muller, \& Ruffert 1998).
However there is no sign of fission into separate condensations.  The
bar evolves, producing two spiral arms that transport the core's
excess angular momentum outwards (see also Shibata, Baumgarte \& Shapiro
2000). This reduces the bar's angular momentum such that a single
central body is formed. How can a body develop local condensations when
bar-unstable ?

According to Bonnell's picture, the evolution of the bar instability 
is more complex, in reality.  If the rapidly
spinning proto-neutron star core goes bar unstable when surrounded by a
fall-back disk, then matter present in the bar-driven spiral arms 
interacts with this material.  The sweeping of a spiral arm into
fall-back gas can gather sufficient matter to condense into a fragment of
neutronized matter. This occurs because the $m=1$ mode grows during the
development of the $m=2$ mode. The $m=1$ mode causes the displacement 
of the unpinned (free-to-move)
core and creates an off-center spiral arm that sweeps up more 
material on one side
than the other during "continuing" accretion. The condensation eventually
collapses into a low mass neutron star.  

Detailed simulations confirming or dismissing the occurrence of such
instability are still lacking, so further considerations of the lump masses,
temperatures and entropy contents are
necessarily speculative.

\subsection {Cooling scenarios and the minimum mass}

Here we wish to explore the possibility that a light
(proto-)neutron star forms around 
the main central body,
lives for a while, and later explodes,   
imprinting a kick to the  neutron star  that remains (due to linear
momentum conservation)  
before gravitational waves or
hydrodynamical effects can drive it toward the central star.  Can a
light (proto)-neutron star form from the condensation of
material
accumulated in the off-centered spiral arm ?
What limits can be imposed on its mass ?

Cooling 
plays a key role in addressing these questions.
Goussard, Haensel \& Zdunik (1998),
Strobel, Schaab \& Weigel (1999), and  Strobel \& Weigel (2001)
have shown that 
the value of the minimum stable mass $m_{\mmc}$ for a neutron star 
(located at the turning 
point in the mass-radius relation of equilibria) 
is a function
of the temperature:
it  varies from $\simgreat
1 \msun$ at 50-100 milliseconds after core bounce (setting the actual
value of the mass of the central neutron star), to $\sim 0.7\msun$ after
$\sim 1$ second, down to $\sim 0.3\msun$ after 30 seconds, reaching the
value of $\sim 0.0925\msun$ (Baym, Pethick \& Sutherland 1971)
known for cold catalyzed matter ($T<1$ MeV)
several hundred seconds later.
Thus, 
the lump of nuclear matter gathered in the spiral arm by the instability
may become self-bound if its mass $m$  is above the minimum corresponding
to that particular temperature.
Once formed it is stabilized against expansion by cooling.
The actual value of $m$ is thus determined by the overall dynamics of
collapse after core bounce and varies between 0.0925 and $\sim 1\msun:$
It depends on the time at which the instability sets in, on the amount
of fall-back material (potential reservoir of matter in the lump)
and on the cooling history.

The value of $m$ and  of the mass ratio $q=m/M$ in the binary 
(with $M$ the heavier of the two 
stars) 
remains  unpredictable to us at this level,  so  
we may just depict three possible scenarios for the formation and evolution 
of this evanescent binary: case (A) when the instability sets after
few hundreds 
of milliseconds or a second after core bounce and 
binary formation occurs so that 
$m\simgreat 0.7\msun$ (implying a
pre-existing iron core of  large mass if the primary is as massive as
1.4$\msun$); 
case (B) when a lighter star forms
around a main body with $m\simeq 0.2-0.4\msun$
after several tens of seconds from core bounce; 
case (C) when cooling is
sufficiently advanced that the minimum mass approaches its asymptotic
value and the binary  can have 
$m\simless 0.2\msun$.

The magnitude of a natal kick in (A) is difficult to estimate and 
we must await for realistic simulations of core collapse.  Values
of $m/M$ above a given threshold 
 in a binary are known to lead to unstable mass transfer 
and thus to final coalescence.  We note
that the evolution in this case might be similar to that described in
coalescing neutron star binaries with slightly unequal masses (see Rosswog
et al. 2000) where it has been shown that large kicks can be acquired
as a consequence of mass loss via a wind.
Case (B) and (C) will be explored in this paper.  
Imshennik \& Popov (1998) estimated neutron star recoil speeds of order
$1500-2000\kms$ for scenario (C).

In cases (B) and (C), one factor that could lower the final neutron star
speed is the finite velocity of the ejecta.  Colpi, Shapiro \& Teukolsky
(1993) have shown that in the dynamical phase of the explosion the
ejecta can attain speeds varying from 10,000 to $\sim 50,000 \kms$ (the upper
bound being related to the value of the binding energy of the star at
the minimum mass relative to a dispersed state of iron). These speeds
are close to the escape velocity from the binary and therefore the final
kick imparted to the remaining neutron star may be influenced
substantially by gravitational deflection of the ejecta,
and
velocity phase averaging during the explosion. A further
threat would be a rapid decay of the orbital separation due  
to unstable mass transfer that may lead to final coalescence, 
and emission of gravitational waves.

In $\S 3$ we
will describe the gravitational bending of the ejecta, while 
in $\S 4$ we
will study initial conditions and, subsequently, 
orbit decay and mass exchange, for cases (B) and
(C).  We will then give an estimate of the kick speed
including phase-velocity averaging.

\section{Gravitational bending of the exploding debris and the final kick}
\label{bending}

Consider now the instant at which the light secondary of mass $m$
explodes, having reached the dynamical instability point ($m\simeq
m_{\mmc}$), while orbiting the heavier, primary neutron star of mass
$M$.

Obtaining the final kick speed imparted to the heavy neutron star is
complicated by several factors. First, some of the material ejected from
the exploding star can remain bound to the system and eventually
accretes onto the remaining star. Second, material that escapes may not
be moving at extremely large speeds, even right at ejection, and
therefore emerges with a different momentum than it has at the point of
explosion. Third, the mass of ejected material may be of order 10\% of
the total mass of the system, and self gravity could play a role.
Fourth, the ejecta are a fluid, and, depending on how much time elapses
between neutron star formation and the explosion of the low mass star,
move through ambient gas left over from the original supernova
explosion.  If the ejecta are slowed significantly by the surrounding
gas, some material that would be judged unbound upon ejection might
actually end up falling back toward the neutron star, possibly to
accrete onto it.

Suppose that relative to the remaining heavy neutron star, the orbital
velocity of the exploding star is $\Vvec$ at the instant of explosion
($t=0$), and the position of the center of mass of the exploding star is
at $\rvec.$ Let us write the final velocity of the remaining neutron
star as \be \Vvec_{\rm{kick}} =-\etavec {mV\over (M+m)},
\label{Pkick}
\ee 
where $V=\vert\Vvec\vert$. If the ejecta
are expelled isotropically relative to $\rvec$, and at speeds large
compared with $V$, then we expect $\eta=\vert\etavec\vert\simeq 1$,
in which case the remaining neutron star recoils with the maximum
possible kick speed
\be 
V_{\rm {kick,max}} \simeq {mV\over (m+M)}
\label{Vkick}
\ee 
as implied by momentum conservation in the center of mass of the  binary.

To get a more realistic, but still approximate, idea of the size of
$\eta=V_{\rm {kick}}/V_{\rm {kick,max}}$, we work to zeroth order in the
ejected mass, and also treat the trajectories of the ejecta
ballistically; this is equivalent to ignoring the third and fourth
complications listed above entirely. By working to zeroth order in the
mass of the ejecta, we can also assume that all particles are ejected
virtually from a single point, $\rvec$.  Let the velocity of a particle
relative to the exploding star be $\wvec$, and assume that $\wvec$ is
distributed isotropically about the position of the exploding star with
probability $P(\wvec)$. Since we work to zeroth order in the mass of the
ejecta, we may take the heavy neutron star to remain at fixed position
in the calculation.

The problem that remains is the orbital mechanics for each
particle relative to the massive star. The orbits are characterized by the conserved 
quantities
\baray
E&=&{1\over 2}\vert\Vvec+\wvec\vert^2-{GM\over r}\nonumber\\
\Jvec&=&\rvec\crossprod(\Vvec+\wvec)\nonumber\\
\Avec&=&-{GM\rvec\over r}+(\Vvec+\wvec)\crossprod\Jvec,
\label{conserved}
\earay
which are the orbital energy and angular momentum per unit mass,
and the Runge-Lenz vector, respectively. We note that
\be
A=\vert\Avec\vert=GMe=\sqrt{(GM)^2+2EJ^2};
\ee
$e<1$ for bound orbits,
$e>1$ for unbound orbits. 

To calculate $\eta$, we need to find the momentum per unit mass
carried away by the ejecta that escape to infinite distance. 
First, we need to identify the initial velocities of particles that
escape. For these, we must have $E>0$, and from  equation (\ref{conserved})
we find the condition
\be
\Vvec\dotprod\wvec=Vw\mu>{GM\over r}-{(V^2+w^2)\over 2}
\label{escapecond}
\ee
where $V=\vert\Vvec\vert$ and $w=\vert\wvec\vert$, and $\mu=
\Vvec\dotprod\wvec/Vw$. Second, we need to find the mapping 
between initial and final velocity. 
To this aim, it is useful to define a coordinate system in which 
$\ehatthree=\Jhat,\,
\ehatone=\Ahat,$ and 
$\ehattwo=\Jhat\crossprod\Ahat,$
%\baray
%\ehatthree&=&\Jhat\nonumber\\
%\ehatone&=&\Ahat\nonumber\\
%\ehattwo&=&\Jhat\crossprod\Ahat, 
%\earay
where $\Jhat=\Jvec/J$ and $\Ahat=\Avec/A$.
The orbit is then confined to the $1-2$ plane, and we can read
off the final velocity in this system from equation (15.12) in
Landau and Lifshitz, {\it Mechanics}:
\be
\vvec_\infty=[-\Avec\sqrt{2E}+(2E/GM)\Jvec\crossprod\Avec]
\label{vfinal}.
\ee
We wish to express the final velocity $\vvec_{\infty}$ in a coordinate system fixed
in the binary at the point of explosion. We thus choose, explicitly
\be 
\rvec=r\xhat \qquad \rm {and}\qquad\Vvec=V\yhat
\ee
so that the orbital angular momentum of the light star relative to $M$ 
just prior to explosion points along
$\zhat$.  We have explicitly selected the point of explosion to occur
at either peri- or apocenter of the orbit, and for simplicity, we specialize
to circular orbits below. (Both tidal effects and gravitational 
radiation will tend to  circularize the orbits, but, even for eccentric
orbits, we expect tidal  disruption to be likeliest at pericenter.) 
In this coordinate system, 
\be
\Jvec=-\yhat rw_z+\zhat r(V+w_y)
\label{J}
\ee
\baray
\Avec=\xhat[-GM+r(V+w_y)^2+rw_z^2]
\nonumber\\+\yhat[-rw_x(V+w_y)]
+\zhat(-rw_xw_z)
\label {A}
\earay
and 
\be
E=-{GM\over r}+{w_x^2+(V+w_y)^2+w_z^2\over 2}.
\label{E}
\ee
The velocity of the ejecta can thus be expressed in terms of
the known explosion parameters ($\wvec,V$, and $r$) 
by combining equation (6) with equations (7)-(10).
The result is that the total momentum carried off per unit mass of ejecta is
\be
\uvec=\int_{E(\wvec)>0}{d^3\wvec P(\wvec)\vvec_\infty(\wvec)},
\ee
so that  
$\eta=\vert\uvec\vert/V,$  to zeroth order in $m/M.$ 
As $P(\wvec)=P(w)$ is isotropic we can  separate out the integrals 
\be
\uvec=\int{dw\,w^2 P(w)} \int{d^2\What\vvec_\infty(w\What)},
\ee
and  consider different distributions of ejection
speeds separately. The integration only extends over unbound orbits.
For $w>\sqrt{2GM/r}+V$, all ejecta escape, and for $w<\sqrt{2GM/r}-V$,
no ejecta can escape.

Because of reflection symmetry with respect to the binary orbital plane,
$u_z$ vanishes identically.
The kick is thus given in the orbital plane of the pre-explosion binary.
It is easy to show that when the particles are ejected with equal velocity
$w_o$ so that 
$P(w)=\delta(w-w_o)/(4\pi w_o),$ 
the integrals 
yield $u_x\simeq 
-(GM/r)w_o^{-1}$ and $u_y\simeq V$ for very large values of $w_o$ ($\gg V$)
which is consistent with the requirement that at large ejection speeds,
the outflow reaches infinity with an average velocity comparable to that
of the exploding star.
Figure 1 shows $\eta$ as a function of the dimensionless 
explosion velocity $w_o/\sqrt{GM/r}.$
Notice that there is no kick  (i.e., $\eta$ vanishes) when 
$E=0$ (see eq. [5]).  This occurs at a critical value of the expansion speed
$w_o/\sqrt{GM/r}=\sqrt{2}-V/\sqrt{GM/r}$ which is $\simeq 0.414$ 
for $V\sim\sqrt{GM/r}$ to zeroth order in
$m/M.$ When $E\leq 0$, the mass outflow
vanishes identically and the ejecta fall back to the remaining
neutron star transferring all their  linear momentum to it.

\def\mmc{{\rm{mmc}}}
\def\min{{\rm min}}
\def\stab{{\rm stab}}
\def\km{{\rm\, km}}
\def\kick{{\rm kick}}
\def\exp{{\rm exp}}

\section{Final Kick Speed from the Short-Lived Binary}

\subsection{Masses and orbital separations}

The final kick speed imparted to the remaining neutron star depends
crucially on the evolutionary scenario leading to the formation and disruption of
its low mass companion.
The range of possible values of $m_{\rm in},$  the initial mass
of the secondary (hereon), depends on the  time the 
fragment forms and how fast its cools.
Both of these complications distinguish the problem of ``early''
formation and decay of an evanescent proto-neutron star binary (case B) from
the problem of ``late'' decay of a neutron star binary (case C)
originally discussed by Clark \& Eardley (1977), Blinnikov et al. (1984),
and Imshennik \& Popov (1998); the final outcome may differ
from the results found by the authors  for
cold neutron stars.

Figure 2 shows an estimate of the binary separation $r_{\rm orb}$ at the 
time of formation  (solid line) as a function of $m_{\rm in}$, which
we regard as an independent variable here. To obtain
this estimate, we assumed that the angular momentum $J_{\beta,\,\rm dyn}$ of
the bar-unstable proto-neutron star with $T_{\rm rot}/\vert W\vert 
\sim \beta _{\rm dyn}$ 
goes entirely into {\it orbital}
angular momentum of the binary. 
(Even if the lump forms in corotation, its angular momentum is smaller
than the orbital angular momentum of the system, substantially so if
the orbital separation is considerably larger than its radius.)
As a reference value  
for $J_{\beta,\,\rm dyn}$ we adopt 
$\sim 4\times 10^{49}\rm { g\,cm^2\,s^{-1}}$,
the value given by Saijo et al. (2000) for  
differentially rotating compact stars at $\beta_{\rm dyn}$ (note that
this is the minimum $J_{\beta}$ since the 
$m=2$ and $m=1$ modes can grow at values of $T_{\rm rot}/\vert W\vert$
larger than  $\beta_{\rm dyn}$). 
As shown in the Figure, lighter
secondaries can be 
accommodated on wider orbits where the timescale of orbit decay for
emission of
gravitational waves 
$\tau_{\rm GW}=5c^5r_{\rm orb}^4/(256 G^3M_{\rm tot}\,M\,m_{\rm in})$ 
is longer ($M_{\rm
tot}=M+m_{\rm in}$ 
is the total
mass of the collapsing core).

Along the relation $r_{\rm orb}-m_{\rm in}$ 
for $J_{\rm orb}=J_{\beta, \,\rm {dyn}}$, 
the bold dots show
values of $m_{\mmc}$ at different  
cooling times $\tau_{\rm cool}$ according to the cooling models of Strobel et al. (1999).
The dot-dashed lines in the sample plane correspond to loci
of fixed values of the gravitational radiation decay timescale
for the orbit, $\tau_{\rm GW}$ (from the top, 1, 30, 100 and
300 seconds, as labeled in the Figure).

Notice that gravitational radiation timescales are generally longer
than cooling timescales, with the only exception occurring
when $m_\mmc\sim
0.7\msun.$ 
Coalescence of the two stars is likely to occur in this case
with some mass loss that can only be estimated via hydrodynamical simulations.
In the opposite case when $m_{\rm in}$ and $r_{\rm orb}$ are such
that $\tau_{\rm GW}$ is longer than the cooling time $\tau_{\rm cool},$ 
the binary survives for a while and below we study its evolution.

\subsection{Binary evolution, mass transfer, and maximum kicks}

In the case of late formation (case C), 
a phase of stable mass exchange may
set in, driven by gravitational radiation emission, which
terminates when the secondary reaches a critical mass, 
$m_\tid$ (Clark \& Eardley 1977; Blinnikov et al. 1984;
Jaranowski \& Krolak 1992; Bildsten \& Cutler 1992; Imshennik \& Popov
1998). Subsequently, the low mass star (which has a core-envelope
structure) loses its extended halo 
evolving along a sequence of hydrostatic equilibria
while the orbit widens (Blinnikov et al 1984); the star reaches the minimum
stable neutron star mass, whereupon it explodes (Page 1982; Colpi,
Shapiro \& Teukolsky 1989, 1991, 1993; Blinnikov et al. 1990;
Sumiyoshi et al 1998).  
In the case of early formation of the binary (B) we show below that 
mass transfer is likely to be unstable, and we argue that 
because of this it may just drive the star to the point of
exploding dynamically.

Consider  a binary in which
Roche lobe overflow is in effect; then, the radius of the low mass
star $R(m)$ overfills the Roche lobe and this
determines the typical orbital separation at the onset of mass transfer
\be
r=2.2R(m)(M+m)^{1/3}/m^{1/3}.
\ee
The dotted lines in Figure 2 give the tidal radius as a function of $m$
for the two different mass-radius relations
describing a cold and warm neutron star, respectively.
A ring/disk of material forms around the primary star at the moment
of Roche filling, and accretion begins to the primary.
As the disk drains  angular momentum from the orbit
there may not be sufficient time
to return it to the orbit through disk-donor tidal torques
(see for details Bildsten \& Cutler 1992; Lubow \& Shu 1975).
If $1-f$ is the fraction of orbital angular momentum stored in the
disk, 
the orbital separation would vary with time
(following Bildsten \& Cutler 1992) as
\be
{\dot {r}\over r}=-{2\dot m(fM^2-m^2)\over Mm(M+m)}-{
64G^3mM(M+m)\over 5 r^4}.
\ee
Fitting the values of $f(m/M)$  computed by Hut \& Paczynski (1984) 
for the Roche problem, Bildsten \& Cutler (1992) found
\be
f\sim (5/3)(m/M)^{1/3}-(3/2)(m/M)^{2/3}.
\ee 
During Roche spillover $m/R(m)^3\propto
(M+m)/r^3$  and  the rate of change of $r$ is 
\be 
{\dot {r}\over r}=\left 
({d\ln R(m)\over d \ln m}-{1\over 3}\right ){\dot m\over m}
\ee
where $\dot m$ is the mass transfer rate.
Using this result in equation (14) for the orbital evolution, we have
\be 
\left [{1\over 3}-{d\ln R(m)\over d \ln m}-{2(fM^2-m^2)\over M(M+m)}
\right ]{\dot m\over m}={
64G^3mM(M+m)\over 5 r^4}. 
\ee
Mass transfer is stable 
only if 
\be 
{1\over 3}-{d\ln R(m)\over d \ln m}-{2(fM^2-m^2)\over M(M+m)}<0
\ee given the fact that
$\dot m/m<0.$

In case (C), if we adopt the mass-radius relationship as in
Jaranowski \& Krolak (1992) valid for a cold star (and mass
below $1 \msun$),
$R(m)=R_0m^b/(m-m_0)^a$ where $R_0=7.5$ km, $m_0=0.09\msun,$
$b=0.79,$ and $a= 0.83,$ we find that the condition
for stable mass transfer is 
\be
T\equiv {1\over 3}+{a+(a-b)(m/m_0)
\over m/m_0-1}-{2(fM^2-m^2)\over M(M+m)}<  0.
\ee
In general there is only a finite interval for the mass $m$ 
for stability whose boundaries are determined by the two root of the
equation $T=0$; $m^-_\tid$ (the lower bound) and $m^+_\tid$ (the upper
bound).  When $m^-_\tid<m<m^+_\tid$ mass transfer is stable. Outside
this interval stability is lost. There may also be no roots, 
in which case the fate of the light star depends on a number of details that
we will discuss below.

Adopting a constant value for $f,$ and setting $\beta\sim \gamma,$
to simplify analysis, 
we note that one regime where the necessary inequality fails
is at low masses, near $m\simeq m_0,$ 
\be
{m^-_\tid \over m_0}\simeq 1+{\beta\over 2f-1/3}
\ee
(For $f=0.4$ and $\beta=\gamma=0.8$,  $m^-_\tid/m_0=2.7$ while 
if $f=1$, $m^-_\tid/m_0\simeq 1.5.$)
The other regime of instability is at masses well above $m_0,$
in which case $d\ln R(m)/d\ln m$  is relatively small, and $m=m^+_\tid$ with 
\be
{m^+_{\tid}\over M}\simeq {1\over 2}\sqrt{4f-{23\over 36}}
-{1\over 12}.
\ee
(For $f=0.4$,  $m^+_\tid/M\simeq 0.4$ while for $f=1,$ 
$m^+_\tid/M\simeq 0.8.$)
Thus, if there is a disk that stores orbital
angular momentum (i.e., $f<1$), 
the range of masses for which there can be stable,
Roche filling transfer shrinks considerably relative to the case $f=1$
explored in Blinnikov et al. (1984)  and Imshennik \& Popov (1998).

We have computed the roots of equation (19) numerically for different $R(m)$.
For the Jaranowski \& Krolak (1992) parametrization of $R(m)$ for
cold neutron stars (C) we find 
the following intervals of $(m^-_\tid,m^+_\tid)$ in units of $\msun$:
(0.14,1.08) or (0.14,1.33) for $f=1$ and $M=1.4\msun$ or $1.7
\msun$;
for $f$ given by equation (15), there are no roots for $M=1.4\msun$, and 
the roots are (0.29,0.50) for  $M=1.7\msun$;
the stability interval
$(m^-_\tid,m^+_\tid)$ depends sensitively
on the logarithmic derivative of
$R(m).$ Note that the root $m^-_\tid$ exists because of
the special dependence of $R(m)$ on the parameter $m_0.$
Numerical hydrodynamical models for the description of 
these semi-detached tight binaries
are necessary to address the problem of mass transfer, 
and preliminary results points toward the existence 
of stable phases (Davies, private communication).
Below, we will adopt the values of $m^-_\tid$ and
$m^+_\tid$ for a 1.7$\msun$
neutron star.

What is the fate of the binary then ?
In case (C)
if the two roots $m^-_\tid$ and $m^+_\tid$
exist, and if the initial mass of the low mass star is 
$m^-_{\tid}< m_{\rm in}<m^+_\tid(\ll M)$, the binary may first go
through a stage of evolution via gravitational radiation without mass
exchange, then a phase of evolution with mass exchange at a rate
determined by the rate of angular momentum loss due to gravitational
radiation, and finally become unstable when $m^-_\tid$
is reached (Blinnikov et al. 1984; Imshennik \& Popov 1998). If,
on the other hand, $m_\mmc\leq m_{\rm in}\leq m^-_\tid$, 
then the binary tightens
as a consequence of gravitational radiation without mass loss, 
until $m$ fills its Roche lobe, whereupon it becomes unstable.
If we define $m_\stab$ to be the  stellar mass at the onset of instability,
then $m_\stab=m^-_\tid$ if $m_{\rm in}>m^-_\tid$,
and $m_\stab=m_{\rm in}$ if
$m_\mmc\leq m_{\rm in}\leq m^-_\tid$.

What happens once $m=m^-_\tid$ sets in has been described by Blinnikov
et al. (1984). As already mentioned, they argue that although the 
equilibrium radius of
a low mass neutron star may become larger than the Roche radius, the vast
majority of its mass may still be enclosed within it (99$\%$
is contained inside the inner 38 km or so). The stripping
of the stellar mass envelope occurs slowly enough that the companion
evolves through a series of nearly equilibrium 
states until $m_\mmc$ is attained, while the orbit widens. 

More accurately, equation (14), in the absence of  gravitational wave
losses, 
leads to a change in orbital radius $r$ as a function of mass loss equal to
\be
{d \ln r \over d \ln m}= -2f + {2fm\over M+m}+ {2m^2\over M(M+m)}
\ee
which can be integrated to yield  when the mass is $m$ 
\be
{r\over r_i}\sim \left ( {m_{\rm {in}}\over m}  \right )^{2f}
\exp\left [ {2(1-f)(m_{\rm {in}}-m)\over M+m}\right ],
\ee
neglecting changes in $M$.
In this approximation the orbit widens by a factor 
$\sim (m_\stab/m_\mmc)^{2f}$ as the mass decreases from 
the  value $m_\stab$ 
toward $m_\mmc.$

The orbital radius at the point of
explosion is somewhat larger than  the radius
of Roche spillover, 
and is 
accordingly
\be
r_\mmc\simeq {2.2 R(m_\stab)M^{1/3}m_\stab^{(2f-1/3)}\over
m_\mmc^2}.
\ee
The orbital speed of the low mass neutron star when it explodes
is 
\be
V\simeq\sqrt{GM\over r_\mmc}
\simeq{0.68G^{1/2}M^{1/3}m_\mmc\over m_\stab^{(f-1/6)}[ R(m_\stab)]^{1/2}},
\ee
implying a maximum kick speed 
\be  
V_{\rm kick,max}\simeq {m_\mmc \over (m_\mmc+M)}\,V\, 
\simeq  {0.68G^{1/2}m^2_\mmc\over 
m_\stab^{(f-1/6)}[ R(m_\stab)]^{1/2}}M^{-2/3}
\ee
that scales as $M^{-2/3},$ due to the dependence of $V$  on $r_\mmc$. This
is a direct consequence of  the hypothesis of Roche lobe mass transfer; 
the system has lost
memory of the initial orbital separation.
Also note that in $V_{\rm kick,max}$ we have
$m_{\rm mmc}$ at numerator, even when $m_\stab=m^-_\tid.$
When $m_\stab=m^-_\tid\sim 0.29 \msun,$ (or when $m_\stab=m_\mmc=0.0925\msun$) 
the maximum kick is $V_{\rm kick,max}\sim 936 \kms$ ($800\kms$), and 
$f\sim 0.4$.

Concerning case (B), warmer proto-neutron stars  seem  to satisfy a smoother
mass-radius relation.  A fit to Strobel et al. (1999)
data  gives   $R(m)\propto m^{-6/5}$ 
which would lead to a stability 
interval
$(m_\mmc, m^+_\tid)$ only for $f$ close to unity (note
that there is only a single root in this case). Since
disk-donor torquing may act on a timescale longer
than the lifetime of the binary, 
it is likely that mass transfer is always unstable under
these circumstances, and as soon as the light star  
overfills its Roche lobe at $m_{\rm in}$, unstable mass transfer begins
and there might be time for a 
phase of common envelope evolution.  In this context, we wish to argue that
the star is driven  almost immediately
toward explosion.
The cooling time $\tau _{\rm cool}\sim 50$ sec (case B; see Figure 2)  
 exceeds the typical time for mass transfer 
$\tau_{\rm mass}\sim m_{\mmc}/{\bar\rho}c_{\rm sound}R^2 \sim 0.02 $
sec  
($\bar \rho\sim 10^{11}\gcm3$ denotes
the density, close to neutron drip, of the mass losing star 
in its envelope,
$R\sim 50$  km the size of the Roche lobe
and $c_{\rm sound}$ is the sound speed in units of $10^{10}\kms$).
In this case,
the mass losing star encounters the instability point $m_\mmc(T)$ 
before being stabilized
by cooling.  We speculate that the core
of the light star explodes on its own  dynamical time
($\sim 0.001$ sec) before
coalescence of the two stars in completed over a orbital
period of $P_{\rm orb}=2\pi (r^3_{\mmc}/GM)^{3/2}\sim 0.03$ sec.
Common evolution is known to be  accompanied by ejection of
part of the envelope enshrouding the system, and this mass loss
may provide an additional thrust to the merged object.
Only hydrodynamical simulations can give 
credit to this scenario, and we expect 
a continuous transition in the physics when moving
from C (cold late type mode) to A (hot hydrodynamical mode)
passing through B(warm early time mode).
If we were to compute $V_{\rm kick,max}$ by assuming that the
warm star explodes right at the time it fills the Roche
lobe, we would obtain nominal speeds close to $10,000 \kms.$
The actual value of the kick velocity is computed below
combing gravitational bending with orbital phase averaging of the
momentum impulse.

\subsection{Velocity phase-averaging and kick speeds}

Our estimate of $V_{\rm kick}$ for cases (C) and (B) using
the expression for $\eta$ would be not complete 
without considering 
that explosion does  happens non instantaneously.
The star explodes on a timescale comparable to
the dynamical timescale, and 
this can be close to the time of revolution in the binary.
In case (B) in particular this effect might be severe.

The final kick speed will be diminished
if the explosion is not instantaneous to a good approximation. The orbital
frequency when the companion explodes is 
\be
\omega_\mmc=\left ({GM\over r_\mmc^3}\right )^{1/2}
\simeq{0.314G^{1/2}m_\mmc^3\over m_\stab^{(3f-1/2)}[R(m_\stab)]^{3/2}},
\ee
which ranges from $\omega_\mmc\simeq 17$ s$^{-1}$ for $m_\stab
=m_\mmc$ to $\omega_\mmc\simeq 360$ s$^{-1}$ for $m_\stab=
m^-_\tid\simeq 0.29\msun$ and $M=1.7\msun$ (case C), and $185$ s$^{-1}$
for case B with $m_\mmc=0.3\msun$. 
At the time of explosion, the light star overfills 
its Roche radius
$R_{R,\exp}\sim R(m_\stab).$
Matter ejected from the unstable star at a velocity $w_o$ crosses
it in a time $R_{R,\exp }/w_o$ and
we can take $R_{R,\exp }/w_o$ as an estimate of the 
duration of the explosion, $\tau_\exp.$

The explosion is almost instantaneous as long as the
dimensionless combination 
\be
{\omega_\mmc R_{R,\exp}\over w_o}\ll 1.
\ee
For $m_\stab\simeq 0.0952\msun$
the dimensionless ratio is 
$2,800\kms/w_o\sim 0.1$ for $m_\stab\simeq m_\mmc$ (case C),
and $9,000/w_o\kms\sim 0.3$ for $m_\stab=m_\mmc=0.3$ in case B, 
taking  $w_o\sim 30,000
\kms$ as an example.

To be more quantitative, suppose that the star loses mass at a rate $\dot m(t)=m
\Gamma(t)$ once it becomes unstable, with
\be
\int_0^\infty{dt\,\Gamma(t)}=1.
\ee
Consider circular orbits only, and assume that although mass loss may extend
over many orbital periods, the orbit of the exploding star remains unaltered
during the mass loss. (The latter approximation should be valid if mass loss
is rapid enough, to zeroth order in the decreasing secondary mass.) 
Adopt a fixed coordinate system defined at $t=0$, when mass loss begins, so
that $\xhat=\rvec(0)/r$ and $\yhat=\Vvec(0)/V$. The momentum per unit ejected
mass is now
\baray
\uvec&=\int_0^\infty{dt\,\Gamma(t)\left[u_x(\xhat\cos\omega
t+\yhat\sin\omega t)\right ]}+\\
&\int_0^\infty{dt\,\Gamma(t)\left[
+u_y(-\xhat\sin\omega t+\yhat\cos\omega t)\right]}\nonumber\\
&\equiv u_+\ehat_-\Gamma_\omega+u_-\ehat_+\Gamma^\star_\omega,
\nonumber \earay
where $u_\pm=(u_x\pm iu_y)/\sqrt{2}$, $\ehat_\pm=(\xhat\pm i\yhat)/\sqrt{2}$,
and
\be
\Gamma_\omega\equiv\int_0^\infty{dt\,\Gamma(t)e^{i\omega t}}.
\ee
Here, $(u_x,u_y)$ are the same as were computed in \S~\ref{bending} for
instantaneous mass loss (at $t=0$).
When the finite duration of the mass loss is accounted for, the efficiency 
of the recoil is diminished from $\eta$ (as computed in \S~\ref{bending}
for instantaneous mass loss) to $\eta\Gamma_\omega$.
For example, if mass is lost at a constant rate, then $\Gamma(t)=\tau_\exp^{-1}
\Theta(\tau_\exp-t)$, 
and we find
\be
\Gamma_\omega=
{\vert\sin(\omega_\mmc\tau_\exp/2)\vert\over\omega_\mmc\tau_\exp/2};
\ee
if instead $\Gamma(t)=\tau_\exp^{-1}\exp(-t/\tau_\exp)$ then
\be
\Gamma_\omega={1\over\sqrt{1+(\omega_\mmc\tau_\exp)^2}}.
\ee
The correction due to phase averaging, $\Gamma_\omega$,
is only modest (and somewhat model dependent) as long as
$\omega_\mmc\tau_\exp\lesssim 1$, but for long explosions, the efficiency
is diminished by a factor of order $(\omega_\mmc\tau_\exp)^{-1}$ in general.

Figure 3 shows the kick velocity $V_{\rm {kick}}$
of the neutron star that remains (of mass 1.7$\msun$; but for
1.4 $\msun$ scale the velocity with $M^{-2/3}$) 
as a function 
of the speed of the ejecta $w_o$ for the two cases (C) and (B).
The solid curves refer to scenario (C), when $f$ is given by the Roche
value (eq. [15]) and $m_\stab$ is equal to $m^-_\tid=0.29\msun$ (lower curve
before crossover occurring at $w_o\sim 40,000 \kms$)) and $m_\mmc$ (upper curve before crossover). The dotted cure is for 
$m_\stab=m^-_\tid \sim 0.14 \msun.$
The dash-dotted line, for case (B), is computed starting the explosion
during common envelope, when $m_\mmc\sim 0.3\msun$ and when the orbit
separation is $P_{\rm orb}/\tau_{\rm cross}= P_{\rm orb} (V/r) $ times less than
the separation at the moment of filling the Roche lobe (eq. [13]),
to mimic orbit decay during common envelops.
We have included the phase-average orbit correction as given
in equation  (33).

Very high kicks come from such complex hydrodynamical
process that alternative models have not been able to produce.

\section {Discussion}

In this paper, we have elaborated on suggestions that substantial neutron
star recoil can result from the explosion of a low mass neutron star formed
in the aftermath of rotating core collapse (Imshennik \& Popov 1998).
A noteworthy feature of this process is that the final kick speed is
determined by nuclear physics.  The details about
the formation mechanism and hydro-dynamical effects produce
in reality a widespread range in velocities, as the one we see.

We have extended
earlier work by estimating the effect of the finite speed of the material
ejected in the explosion, including both gravitational deflection, phase
averaging, and indirectly taking into account
cooling effects. 
We argue that a range of neutron star recoil speeds could
arise from disruption scenarios with different companion masses. 
The recoil speed 
resulting from the dissolution of an evanescent binary system 
can be around 1000 $\kms$, explaining
the larger values
deduced observationally. It is remarkable that nuclear physics implies a
value in the observational range at all, once such a binary system is
presumed to form, a concordance that lends some support to the idea.

The kicks that result from this mechanism are confined to the orbital
plane of the evanescent neutron star binary. We expect the spin angular
momentum vector of the remnant neutron star
to be nearly, if not perfectly, aligned with the
orbital angular momentum of the binary, and in turn
aligned with the angular momentum vector
of the collapsing iron core. Near alignment would be
consistent with the requirements imposed by the observation of geodetic
precession of B1913+16, where the kick is constrained to lie very
nearly in the plane of progenitor binary, which was most likely
perpendicular to the spins of the spun-up neutron star (i.e. B1913+16)
and its pre-explosion companion star (Wex, Kalogera, \& Kramer 2000).
In contrast, though, X-ray observations of the Vela pulsar have
revealed a jet parallel to its proper motion (Pavlov et al. 2000; 
Helfand, Gotthelf, \& Halpern 2001), and it has been argued that 
the proper motions of both Vela and the Crab pulsar are closely aligned
with their spin axes (Lai, Chernoff, \& Cordes 2001). The kick mechanism
studied here would not be able to account for parallel spin and velocity.
However, we note that the proper motions of both Vela and the Crab correspond
to transverse speeds of $70-141\kms$ and $171\kms$ respectively, using
reasonable estimates of the distances to the pulsars (Lai, 
Chernoff, \& Cordes 2001). For the alignment to be real, the space
velocities of these two systems must lie in the plane of the sky.
The inferred speeds of these two pulsars are then considerably smaller than
the characteristic speed arising from explosion of a low mass, tidally
disrupted companion. We propose that the formation of Vela and
the Crab did not produce an evanescent binary,
and therefore some other mechanism was responsible for their spin-aligned
kicks (such as those explored by Lai, Chernoff, \& Cordes 2001).
We suggest that the larger kick component is due to formation and disruption of
an evanescent binary, and is perpendicular to the
spin axis, and that the smaller kick component is associated with
other less vigorous kicks that tend to align with the rotation axis,
perhaps because of phase averaging (e.g. Spruit \& Phinney 1998; 
Lai, Chernoff, \& Cordes 2001). A superposition of these two classes
of kicks would also be consistent with the requirement of {\it nearly}
but not {\it precisely} spin-perpendicular kicks to account for
the observation of geodetic precession in B1913+16 (Wex, Kalogera, 
\& Kramer 2000). If this idea is correct, then one expects that lower
velocity neutron stars should have their space velocities predominantly
along their spin axes, and high velocity neutron stars should have
space velocities predominantly perpendicular to their spins,
giving origin to two independent (low and high velocity) distributions.
A contamination of low velocity stars (belonging to the low velocity tail
of the high velocity distribution) would come from those explosions
in the evanescent binary where light bending and rotational averaging
have been more important.
This could give rise to a {\it bimodal} distribution of velocities as is
inferred from the observations (Arzoumanian, Chernoff, \& Cordes 2001).

This scenario predicts a clear signature in the neutrino
emission, in the aftermath of core collapse: 
Two neutrino bursts occurring several seconds or minutes after core
bounce, i.e., after the main neutrino burst, should signal (i)
Kelvin contraction of the condensation gathered
by the instability and (ii) explosion (on the dynamical time)
of the light neutron star.  Neutrino
emission between the
two bursts should also come form material that accretes onto the neutron star
that remains. 
Heavy 
$r-$process elements, debris of the explosion
of the light star, should also be ejected and found deep
in the supernova expanding shells. 

Finally, we note that the same phenomenon could occur if the formation
of a light neutron star accompanies rotating core
collapse to a black hole. In that case, we find that the  recoil
velocity of the black hole will be proportional to $M_{\rm BH}^{-2/3}$,
which is not the simple $M_{\rm BH}^{-1}$ scaling one would expect
for a kick mechanism that ejects the same amount of momentum irrespective
of whether a supernova leaves behind a neutron star or a black hole
\footnote {During revision of this manuscript we learned of
a paper by Davies et al. (astro-ph/0204358)
in which a connection bewteen gamma-ray bursts and pulsar 
kicks is made, where kicks result from recoil by a short-lived binary.}.
The higher kick implied by this scaling could be compensated by
stronger relativistic effects of gravitational bending and unstable
mass transfer that can lead to a much lower escape probability of
the exploding debris.

\acknowledgments
The authors thank Melvyn Davies, 
John Miller and Andrea Possenti for useful discussions, and the Referee
for his/her critical comments.
I.W. acknowledges support from NASA Grant NAG 5-8356, and M.C.
from ASI and the Ministry of the University
and Research (MIUR) under Grant MM02C71842-001.

%%%%%%%% figures now

\clearpage

\begin{figure}
\plotone{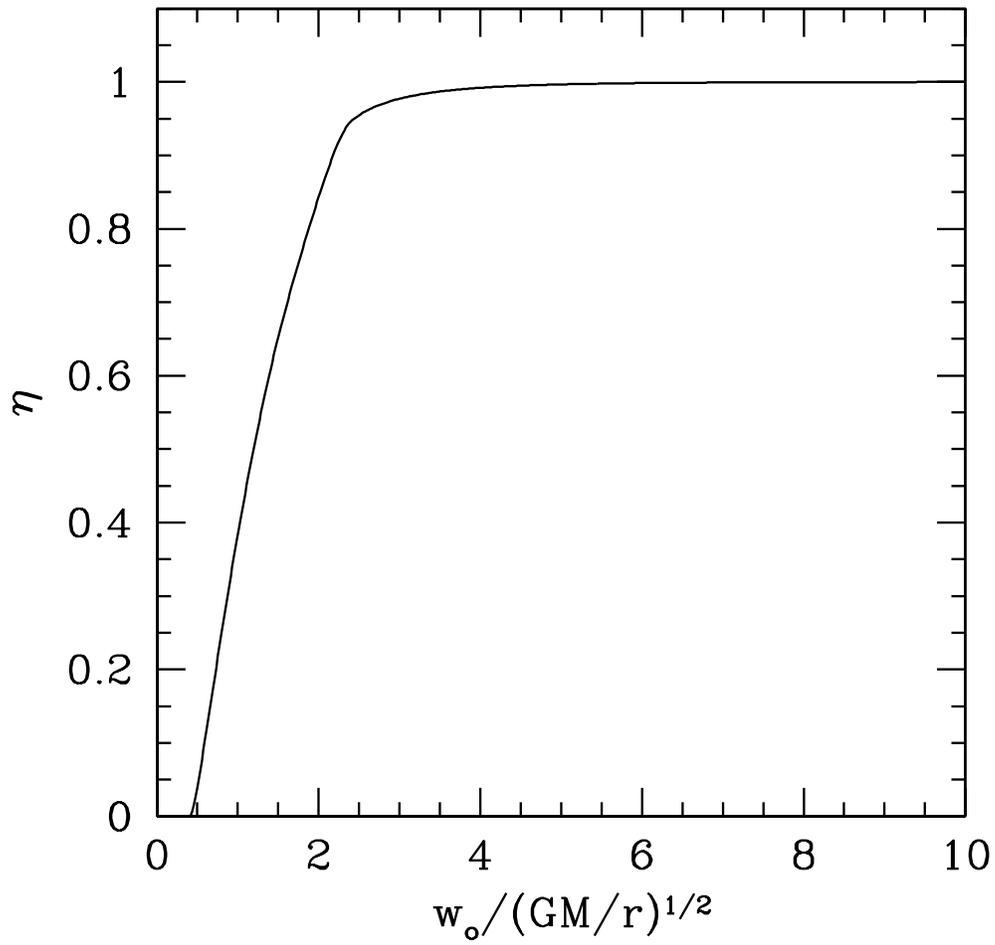}
\caption{The parameter $\eta=V_{\rm kick}/V_{\rm kick,max}$ as a function of 
the  speed of the ejecta $w_o$ expressed in units of $\sqrt{GM/r}.$
\label{fig1}}

\end{figure}
\clearpage

\begin{figure}
\plotone{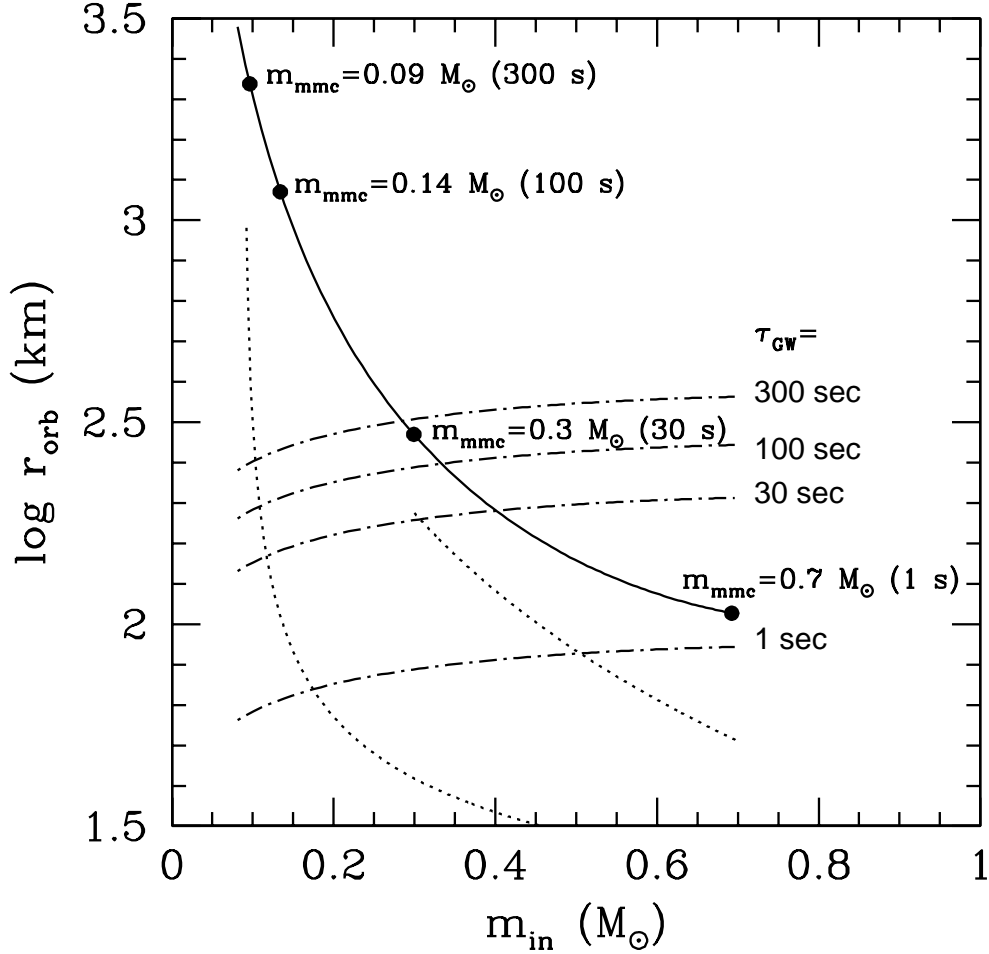}
\caption{Solid line gives the orbital separation $r_{\rm orb}$ (in km) versus $m_{\rm in},$
the initial mass  of the light component, in a binary of total mass
$M_{\rm tot}=1.7\msun$  and orbital angular momentum $J_{\beta,\,\rm
dyn}=4\times 10^{49}$ g cm $^2$ s$^{-1}.$
The dashed lines are the loci in the $r_{\rm orb},m_{\rm in}$ plane 
of constant time $\tau_{\rm GW}$ for
1,30,100,300 seconds. The filled dots denote the various values of $m_{\mmc}$
at the times $t_*=1,20,100,300$ seconds, according to the cooling
calculation of Strobel et al. (1999). The dotted lines give the binary
separation at the time of Roche spillover for a cold star with
$R(m)=R_0m^b/(m-m_0)^a$  as in Jaranowski \& Krolak, and for 
a warm star with $R(m)$
as given from the fit to Strobel's data: $R(m)=R_*(m/m_*)^{-6/5}$
with $m_*=0.3\msun$ and $R_*=48$ km.
\label{fig2}}

\end{figure}
\clearpage

\begin{figure}
\plotone{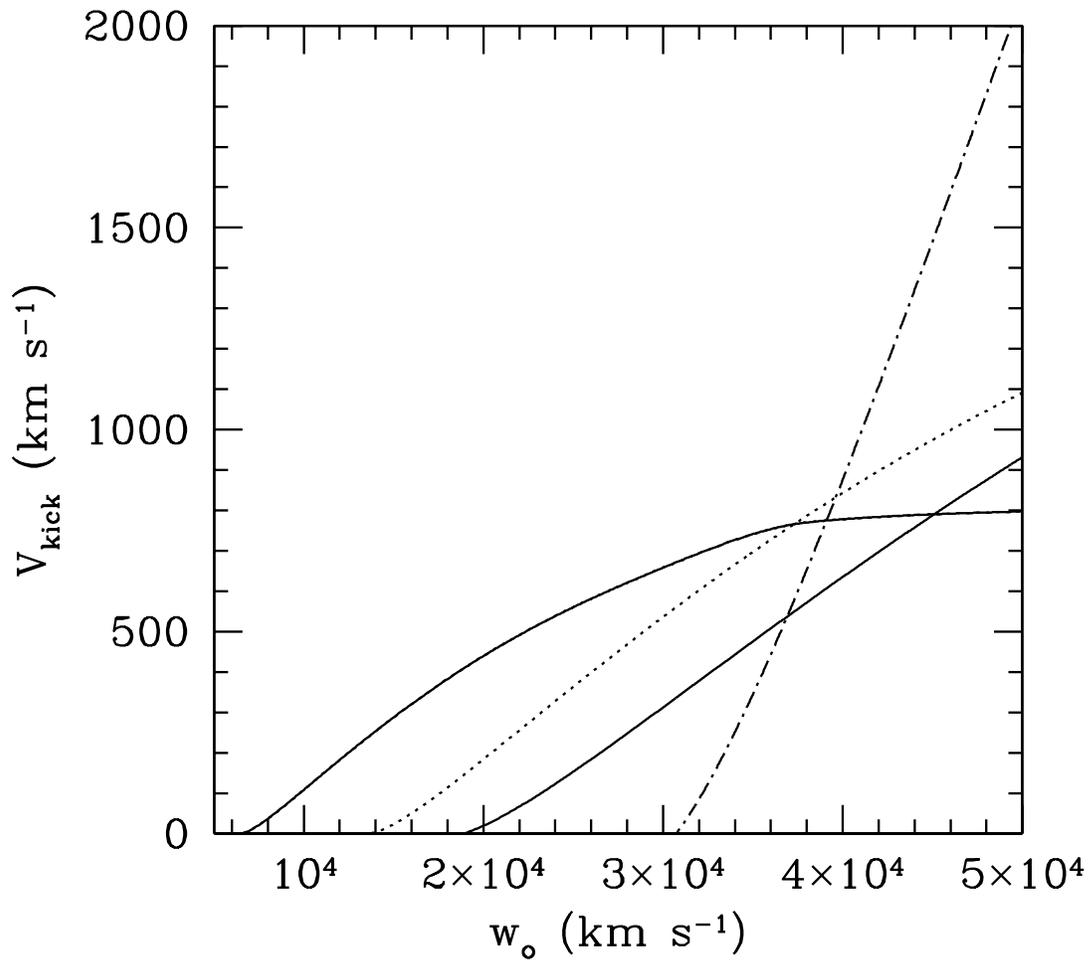}
\caption{Solid lines show the neutron star kick velocity $V_{\rm kick}$ as
a function of the speed of the ejecta $w_o,$ for case (C) when 
$m_{\rm stab}=m^-_\tid$ (lower curve before crossover) and $m_{\rm stab}=m_\mmc$
(upper curve before crossover). The dotted line 
is for $m_{\rm stab}=m^-_{\tid}=0.14\msun$ for $f=1.$  
The dash-dotted line refers to case (B) for $m_\mmc=0.3\msun;$ 
for this case velocity phase-averaging is an important correction.
\label{fig3}}
\end{figure}

\end{document}